\documentstyle[emulateapj]{article}

\newcommand\Ha{H$\alpha$}

\newcommand\hii{\ion{H}{2}}

\newcommand\etal{et\thinspace al.~}

\newcommand\ergs{{\rm\,erg\,s^{-1}}}
\newcommand\kms{{\rm\,km\,s^{-1}}}
\newcommand\vcirc{v_{\rm circ}}

\newcommand\hlf{\hbox{H{\thinspace}{\small II}} LF}
\newcommand\lsat{L_{\rm sat}}

\newcommand\rsat{R_{\rm sat}}
\newcommand\rco{r_{\rm co}}
\received{16 January 2003}
\accepted{}

\slugcomment{Accepted to the {\bf ASTRONOMICAL JOURNAL} 16 June 2003}

\lefthead{Oey, Parker, Mikles, and Zhang}
\righthead{\hii\ region sizes, LF, and corotation}

\begin{document}

\title{	
H\,{\small II} regions in spiral galaxies: \break
Size distribution, luminosity function, and new isochrone diagnostics of
density wave kinematics}

\author{M. S. Oey}
\affil{Lowell Observatory, 1400 W. Mars Hill Rd., 
	Flagstaff, AZ   86001}
\author{Jeffrey S. Parker\altaffilmark{1,2}}
\affil{Whitman College, Walla Walla, WA   99362}
\author{Valerie J. Mikles\altaffilmark{1,3}}
\affil{Johns Hopkins University, Dept. of Physics and Astronomy, 
	Baltimore, MD   21218}
\author{\smallskip and \\ Xiaolei Zhang}
\affil{Naval Research Lab, Remote Sensing Division,
	4555 Overlook Ave., SW, Washington, DC 20375}
\altaffiltext{1}{Participant in the 1999 STScI Summer Student Program.}
\altaffiltext{2}{Present address:  University of Colorado, Dept. of Aerospace,
	Engineering Sciences, Boulder, CO   80309}
\altaffiltext{2}{Present address:  University of Florida, Dept. of Astronomy,
	P.O. Box 112055, Gainesville, FL   32611}

\begin{abstract}

We investigate the relationship of the \hii\ region luminosity
function (\hlf) to the \hii\ region size distribution and density wave
triggering in grand-design spiral galaxies.  We suggest that the
differential nebular size distribution is described by a power law of
slope $\sim -4$, with flattening at radii below $\sim 130$ pc.  This
contrasts with the conventional exponential 
description, but it is physically and quantitatively consistent with the
typical observed value of $-2$ for the \hlf\ slope.

We have developed an interactive code that computes
spatial isochrones for the evolving loci of spiral density waves 
in disk galaxies.  This allows comparison of the nebular spatial
distribution with the spatial isochrones for simple rotation curve
parameters.  Our comparisons for four grand-design galaxies suggest
that the corotation radius $\rco$ coincides with the outer ends of the
star-forming arms.  This value for $\rco$ yields the best spatial
correspondence between the \hii\ regions and the isochrones, and also
appears to yield a coincidence between the Inner Lindblad Resonance
with the radial onset of star formation in the arms.  Thus, {\it we
suggest that isochrones offer a new, simple, and effective technique for
determining $\rco$,} and thus the spiral pattern speed.  However,
application of the isochrones also demonstrates that evolution of the
nebular population is difficult to spatially isolate in these galaxies.

\end{abstract}


\section{Introduction}

The \hii\ region luminosity function (\hlf) is an important diagnostic
of star formation properties in galaxies.  It is a fairly
straightforward parameter to determine for nearby galaxies (e.g., Kennicutt,
Edgar, \& Hodge 1989; Banfi {\etal}1993; Rozas, Beckman, \& Knapen
1996a; Thilker {\etal}2002), and is usually presented in the
differential form as a power law: 
\begin{equation}\label{eqhlf}
N(L)\ dL = A\ L^{-a}\ dL \quad ,
\end{equation}
where $N(L)\ dL$ is the number of nebulae with \Ha\ luminosities in the range
$L$ to $L + dL$.  Occasionally, the cumulative \hlf\ is given, but we
note that the differential form is a more powerful diagnostic of the
underlying properties (Cioffi \& Shull 1991).

In an earlier paper (Oey \& Clarke 1998, hereafter Paper~I), we
modeled the behavior of the \hlf, and identified ways
in which it can be used to examine the current properties of massive
star formation and recent ($\lesssim 10$ Myr) star formation history.
We assumed a power-law distribution in $N_*$, the number of
ionizing stars per object:
\begin{equation}\label{N*}
N(N_*)\ dN_* = B\ N_*^{-\beta}\ dN_* \quad .
\end{equation}
We then performed Monte Carlo simulations, in which we draw $N_*$ from this
distribution, and also draw the individual stellar 
masses from a Salpeter (1955) initial mass function (IMF).
These models showed that reported variations of the measured slope $a$
(equation~\ref{eqhlf}) in different environments {\it do not 
require varying $\beta$}, the slope of the $N_*$ distribution
(equation~\ref{N*}), but rather, can be explained by variations
in the star formation history and the upper cutoff in the $N_*$
distribution.  An important feature is that stochastic
effects for objects with small $N_*$ cause a flattening in the \hlf\
below a certain luminosity $\lsat$.  Sparse clusters subject to this
effect are termed ``unsaturated'' with respect to the IMF, and clusters
with higher $N_*$ we call ``saturated''.  We identified $\lsat$ 
with the \Ha\ luminosity associated with the most
massive star in the IMF, although we refer readers to Paper~I for a
complete discussion.

A basic test on the nature of the \hlf\ is its relation 
to the size distribution of the \hii\ regions.  Although the \hlf\
has empirically been described by a power law, the corresponding size
distribution has been described by an exponential law (van den Bergh
1981; Hodge 1983, 1987).  As discussed above, Paper~I found that the
slope breaks and other features in the \hlf\ are nevertheless best
explained by a simple power law in $N_*$ (equation~\ref{N*} of Paper~I).  
Therefore we should naively expect the size distribution to also
follow a power-law distribution.  In \S 2 below, we examine whether
observations of the \hii\ region size distribution are consistent
with a simple power-law relationship. 

\begin{figure*}
\epsscale{2.0}
\plotone{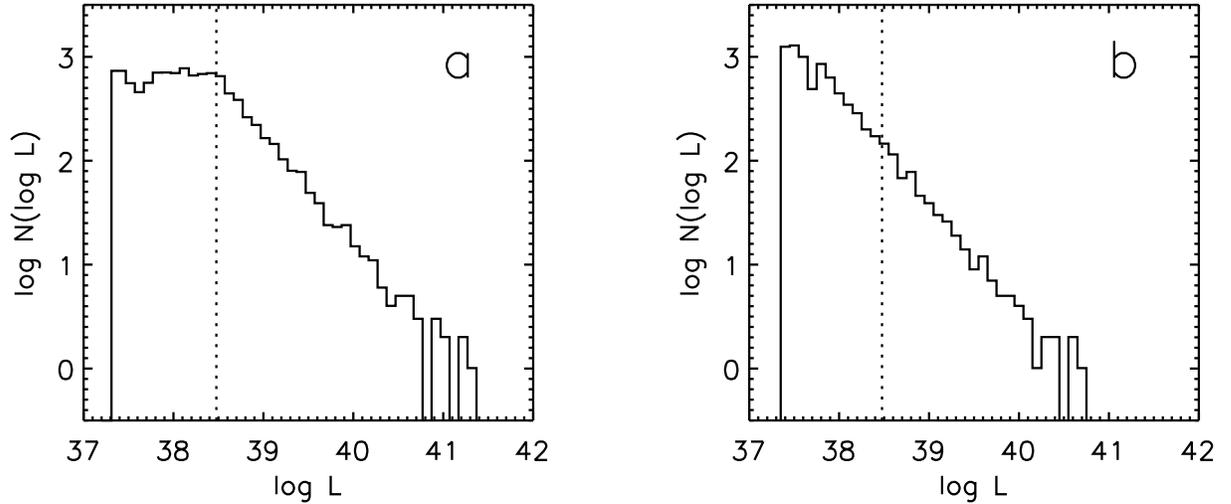}
\vspace*{-5.5truein}
\caption{
Model {\hlf}s: ($a$) zero-age; and ($b$) evolved single-burst, from Oey
\& Clarke (1998; Figures~2$a$ and $c$).  The dotted line marks $\lsat$.
\label{figmod}}
\end{figure*}

Another test on the nature of the \hlf\ addresses its evolutionary
properties.  The models constructed in Paper~I included the evolution
of the \hlf\ resulting from a single burst creation scenario.  In
Figure~\ref{figmod}, we show representative zero-age and 7 Myr models
(Figures~2$a$ and $c$ of Paper~I).  The zero-age model shows the
flattening below $\lsat$.  As time goes by, the entire \hlf\ shifts to
lower $L$.  However, when considering a fixed range of $L$, this can
result in an apparent elimination of the slope break in strongly
evolved populations, due to the differential fading effect.
If this model turns out to be correct, it would then be possible 
to distinguish, for example, between an older
nebular population where high-luminosity objects were created in a
single burst, and continuous creation of low-luminosity \hii\ regions. 

Paper~I applied these results to
observations of arm and interarm regions of grand design spiral
galaxies, suggesting that arm populations represent the current burst
of star formation, while interarm objects represent an aged population
remaining in the wake of the spiral density waves.  Observations
in the literature sometimes suggest that interarm populations show a
slightly steeper \hlf\ than the arm populations (e.g., Kennicutt
{\etal}1989; Rand 1992).  This phenomenon can be explained by
the two-slope structure in the zero-age models, which
will cause a slope measurement to be flatter when fitted over $L$
below $\lsat$ (see Figure~\ref{figmod}).  In contrast, the evolved
populations show the
intrinsic slope of the high-$L$ regime.  Figure~5 of Paper~I shows
that this interpretation of evolution between the arm and interarm
\hii\ regions is consistent, in all cases, with available observations
of six grand design spiral galaxies.

Observations of the \hlf\ in the literature to date have been
subjectively and coarsely binned
into the arm and interarm categories.  A more powerful test of this
evolutionary scenario would be to rebin the \hii\ regions more
quantitatively into regions corresponding to fixed age ranges behind
the leading edge of the spiral arms, as inferred from the spiral arm
pattern speeds.  We might then be able to examine
the competing contributions in interarm regions, of aging nebular
populations versus bona fide interarm star formation.  
These are our motivations for carrying out the studies
described in \S 3.  As an important by-product of this study, we
developed a technique that compares nebular positions with spatial
isochrones for the spiral density waves.  This offers a
new method for locating the corotation resonances in spiral galaxies,
and is described in \S 4.

\section{The form of the \hii\ region size distribution}

As mentioned in the Introduction, the conventional parameterization of
the \hii\ region size distribution as an exponential (e.g., van den
Bergh 1981; Hodge 1983, 1987) is intuitively at odds with the power-law
form of the \hlf.  Since the \hlf's behavior appears to be fairly
well-understood (e.g., Paper~I), it is important to reexamine the
size distribution within the same framework, since they are
physically related.  In particular, we might also expect the size
distribution to follow a power law of the form,
\begin{equation}
N_r(R)\ dR \propto R^{-b}\ dR \quad ,
\end{equation}
where $R$ is the nebular radius.  This has indeed been
suggested for some galaxies (e.g., Pleuss, Heller, \& Fricke 2000).

Since the nebular \Ha\ luminosity $L$ is integrated from the nebular
volume emission, we should expect $L \propto R^3$ and $\frac{dL}{dR}
\propto R^2$.  Therefore, the differential size distribution should be:
\begin{equation}
N_r(R)\ dR = N(L)\ \frac{dL}{dR}\ dR \propto R^{2-3a}\ dR \quad .
\end{equation}
Hence we find that the slopes $a$ and $b$ of the \hlf\ and size
distribution, respectively, are related as $b = 2-3a$.  For the usual value
$a=2$ found for the high-luminosity, ``saturated'' \hii\ regions, we
therefore predict a corresponding size distribution having
$b=4$.  

As mentioned earlier, a turnover in the \hlf\ caused by stochastic
effects for ``unsaturated'' objects occurs at a luminosity $\lsat$.
Since $\lsat$ is associated with the most massive stars in the IMF,
their Str\"omgren radius $\rsat$ should be a corresponding inflection
point in the size distribution.  This flattening at small $R$ can
mimic an exponential distribution.  For the most massive stars,
stellar atmosphere models suggest $\log\lsat/\ergs\sim 38.0$ to 38.5
(Schaerer \& de Koter 1997; Vacca, Garmany, \& Shull 1996; Panagia
1973), implying $\rsat\sim 130$ pc for a density of $1\ \rm cm^{-3}$. 
This estimate for $\rsat$ should be taken as an upper limit,
since the density
is likely to be higher in the actual, central star forming area.

We can now compare observations of the \hii\ region size distribution
to our predictions for slope $b=4$, with a flattening in $R$ below
$\log\rsat/{\rm pc}\sim 2.1$.  We refitted the nebular
diameter distributions with power laws for a number of galaxies
in the literature.  Since size determinations can be sensitive to
different methods, and thus to ensure relatively consistent
comparisons, we used data from only two groups of investigators:  the
IAC--Hertfordshire collaboration (Knapen et al. 1993; Rozas et
al. 1996b; Knapen 1998), and Youngblood \& Hunter (1999).  
Hodge's group (e.g., Hodge {\etal}1994); Miller \& Hodge 1994; Hodge
{\etal}1999) and Petit's group (e.g., Petit 1993, 1996, 1998) have
contributed many studies of \hii\ region size 
distributions; however, their target galaxies generally
do not offer enough objects to adequately sample the nebular size
distribution at $\log D/{\rm pc} > 2.4$, where $D$ is the nebular diameter.
Likewise, we include only the three galaxies from the Youngblood \&
Hunter (1999) sample that provide adequate sampling in this regime.

\begin{figure*}
\vspace*{0.3 truein}
\epsscale{1.0}
\plotone{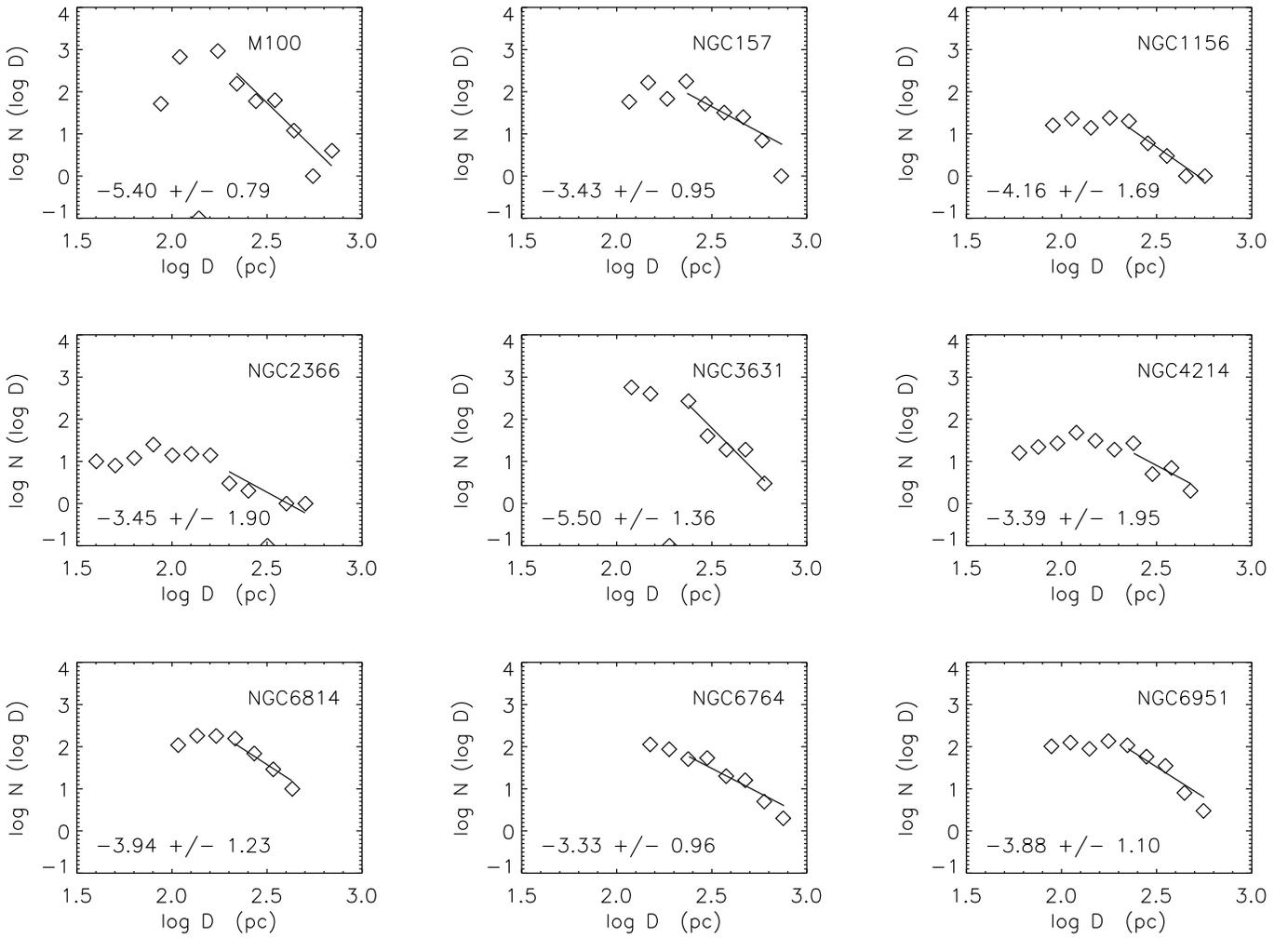}
\caption{Differential \hii\ region diameter distributions for galaxies in
Table~\ref{sizeref}, fitted with power laws for $\log D/{\rm pc} \ge 2.3$.
The fitted values of $-b$ and standard deviation are shown [note that the
fitted slope for $\log N(\log D)$ vs. $\log D$ is $1 - b$\ ].  Empty bins
are shown as $\log N(\log D) = -1$.
\label{sizedist}}
\end{figure*}

Figure~\ref{sizedist} shows the size distributions for the nine
galaxies listed in Table~\ref{sizeref}.  We fitted power law slopes for 
$\log N(D)/{\rm pc} \ge 2.3$, with the bins weighted by
$\sqrt{N}$.  The IAC--Hertfordshire galaxies are grand
design spirals and the Youngblood \& Hunter galaxies are Magellanic
irregulars.  Both these sets of galaxies have \hlf s extending to high
luminosities, consistent with no upper cutoff in luminosity.  Our fits to
the size distributions therefore should not be influenced by possible
truncation at the upper end of the \hlf.  It appears that the \hii\
region diameter distributions shown in Figure~\ref{sizedist} are
consistent with power laws in the given regime.  Objects with smaller
diameters show a flattening in the size distribution,
qualitatively consistent with the flattening in the \hlf\ for
statistically unsaturated objects.

\begin{deluxetable}{lll}
\footnotesize
\tablecaption{References for galaxy diameter distributions \label{sizeref}}
\tablewidth{0pt}
\tablehead{
\colhead{Galaxy} & \colhead{Type\tablenotemark{a}}   & \colhead{Reference}
}
\startdata
M100	& Sc(s)I & Knapen (1998) \\
NGC 157 & Sc(s)I--II & Rozas et al. (1996b) \\
NGC 1156 & SmIV & Youngblood \& Hunter (1999) \\
NGC 2366 & SBmIV--V & Youngblood \& Hunter (1999) \\
NGC 3631 & Sbc(s)II & Rozas et al. (1996b) \\
NGC 4214 & SBmIII & Youngblood \& Hunter (1999) \\
NGC 6814 & Sbc(rs)I--II & Knapen {\etal}(1993) \\
NGC 6764 & SBbc(s)\tablenotemark{b} & Rozas et al. (1996b) \\
NGC 6951 & Sb/SBb(rs)I.3 & Rozas et al. (1996b) \\
\enddata
\tablenotetext{a}{From Sandage \& Tamann 1987.}
\tablenotetext{b}{Type as listed by Rozas et al. (1996b).}
\end{deluxetable}

The small dynamic range and small number statistics cause some
ambiguity and large uncertainties in the fitted slope $-b$, as shown in
Figure~\ref{sizedist}.  We also caution that we cannot rule out the
existence of systematic effects.  Nevertheless, the average slope,
weighted inversely by the uncertainties, is $\langle b \rangle =
4.12$ with a standard deviation of 0.90.  This preliminary result
agrees with the value predicted 
above of $b=4$.  Figure~\ref{sizedist} also shows that the turnover in
the diameter distribution is also consistent with the predicted value
of $\log D/{\rm pc} \sim 2.4$.  {\it We therefore suggest that the \hii\
region size distribution is intrinsically described by a power-law function}
in the regime beyond the turnover, consistent with its physical relation to
the \hlf.  While consistent functional forms are expected for
the size distribution and \hlf, further confirmation of the
empirical power law for the former is desireable, with attention to
systematic uncertainties in size determinations.

\section{Testing for an evolving \hlf}

Based on Paper~I, we approximate that the \hii\ regions found along the
leading edge of spiral arms represent the youngest, zero-age
population of objects.  \hii\ regions behind the leading edge therefore
represent a somewhat older population remaining in the wake of the
spiral density wave.  If the galactic rotation curve and spiral
arm pattern speed are known, it is in principle possible to
estimate the ages of objects based on their location with respect to
the leading edges of the spiral arms.  This would also assume that the
velocities of the \hii\ regions do not deviate significantly
from the galactic rotation curve during the passage of the density wave.  

We have developed a code that computes spatial isochrones from the
rotation curve, given the current locus of the spiral arms.  We defer
discussion of the code and spiral arm kinematics until \S 4 below, and
here use the spatial isochrones to
search for evolutionary features as predicted in Paper~I.  We examined
four of the grand design spiral galaxies studied in Paper~I, while the
two remaining galaxies included in that paper, NGC 3631 and NGC 6814,
did not have a simple two-arm structure suitable for this analysis.
Figure~\ref{isochrones} below shows \Ha\ images of the galaxies:  M51
(Rand 1992), M100 (Knapen {\etal}1993), NGC 157, and NGC 6951 (Rozas
{\etal}1996).  Five sets of spatial isochrones are calculated for each
of the two dominant arms in these galaxies, calculated at intervals of
12 Myr.  The color scheme for times --12, 0, 12, 24, and 36 Myr are:
red, green, yellow, yellow, red for one arm; and blue, yellow, green,
green, blue for the other.  The zero-age isochrones correspond to an
actual spatial fit to the arm positions.  The adopted
corotation radius is shown with the black dashed line, and the inner
and outer Lindblad resonances are shown in the solid and dotted black
lines, respectively, assuming $m=2$ spiral arms for these galaxies.

Unfortunately, it appears that
the isochrone spatial separation that is sensitive to
the nebular evolutionary timescale is of order the spatial scatter of
the objects along the spiral density wave.  Figures~\ref{isochrones}$a
- c$ show that the best-defined arms are enclosed within the
--12 to +12 Myr isochrones, but show a spatial scatter that precludes
further distinction.  NGC 6951 is an exception
(Figure~\ref{isochrones}$d$), showing larger separation between the
isochrones that also corresponds to the nebular spatial distribution.

\begin{figure*}
\vspace*{0.3 truein}
\epsscale{2.0}
\plotone{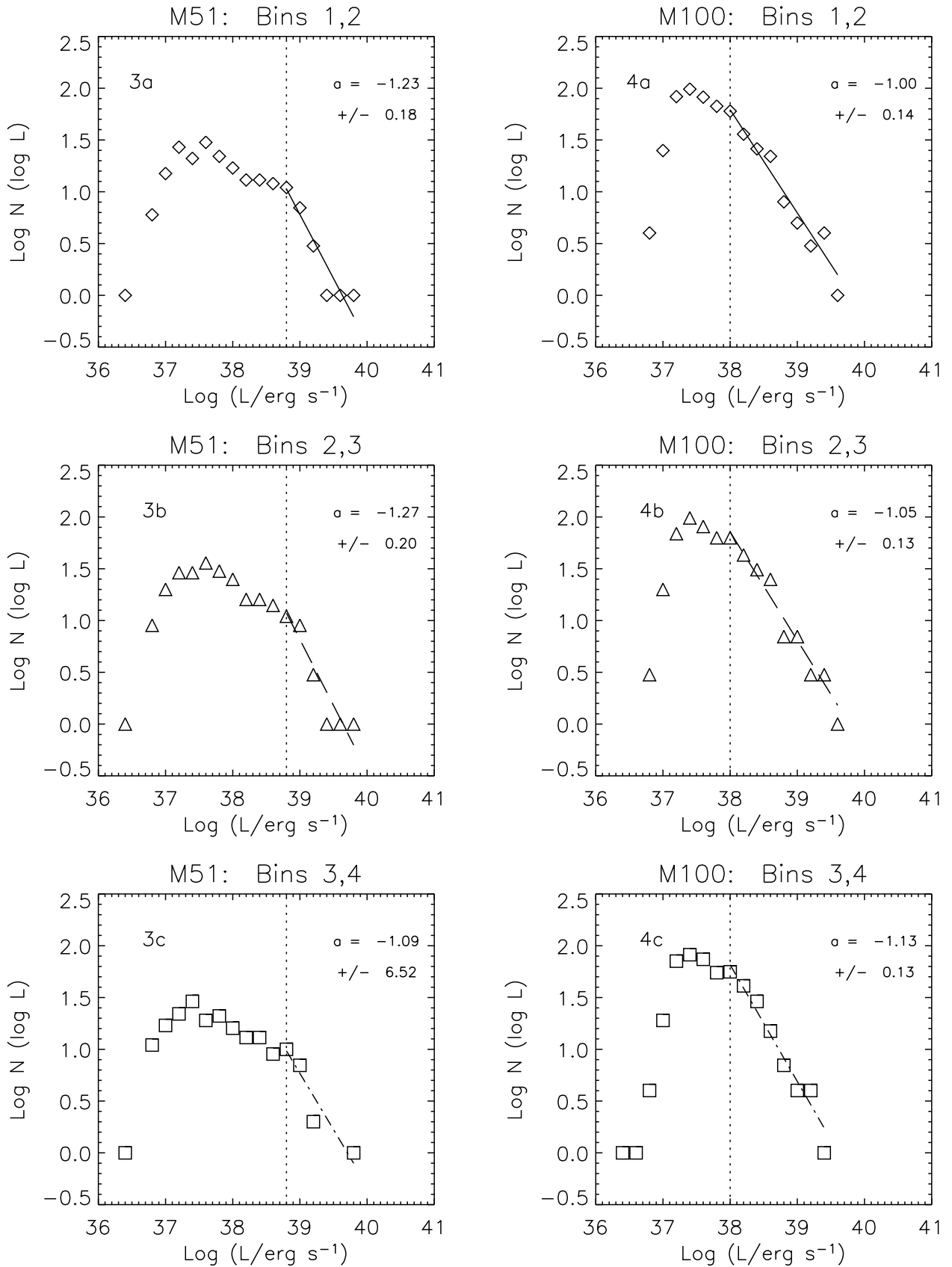}
\caption{6. ---  \hlf s for objects binned between isochrones for M51,
M100, NGC 157, and NGC 6951.
Panels $a, b$ and $c$ respectively show Bins 1 -- 2 (--12 to +12 Myr),
Bins 2 -- 3 (0 -- 24 Myr), and Bins 3 -- 4 (12 -- 36 Myr).
\label{M51hlf}}
\pagebreak
\end{figure*}

Figures~\ref{M51hlf} -- 6 show the \hlf\ for
sub-groups of \hii\ regions binned within isochrones for --12 to +12
Myr (panels $a$); 0 to 24 Myr (panels $b$), and 12 to 36 Myr (panels
$c$).  As discussed above from Paper~I, the \hlf\ naturally 
shows a break in slope below the high-luminosity power-law tail, that
is caused primarily by the transition to unsaturated stellar clusters
dominated by small-number statistics.  For older nebular populations,
this slope transition at $\lsat$ is expected to shift toward lower
luminosities, while the high-luminosity power-law slope remains the
same (Figure~\ref{figmod} above; Paper~I). 
The vertical dotted line in these figures shows the luminosity bin
where we estimate $\lsat$ to occur for the youngest
binning; this value is plotted for reference, and remains
constant between the three \hlf\ subsets for each galaxy.

\begin{figure*}
\vspace*{0.3 truein}
\epsscale{2.0}
\plotone{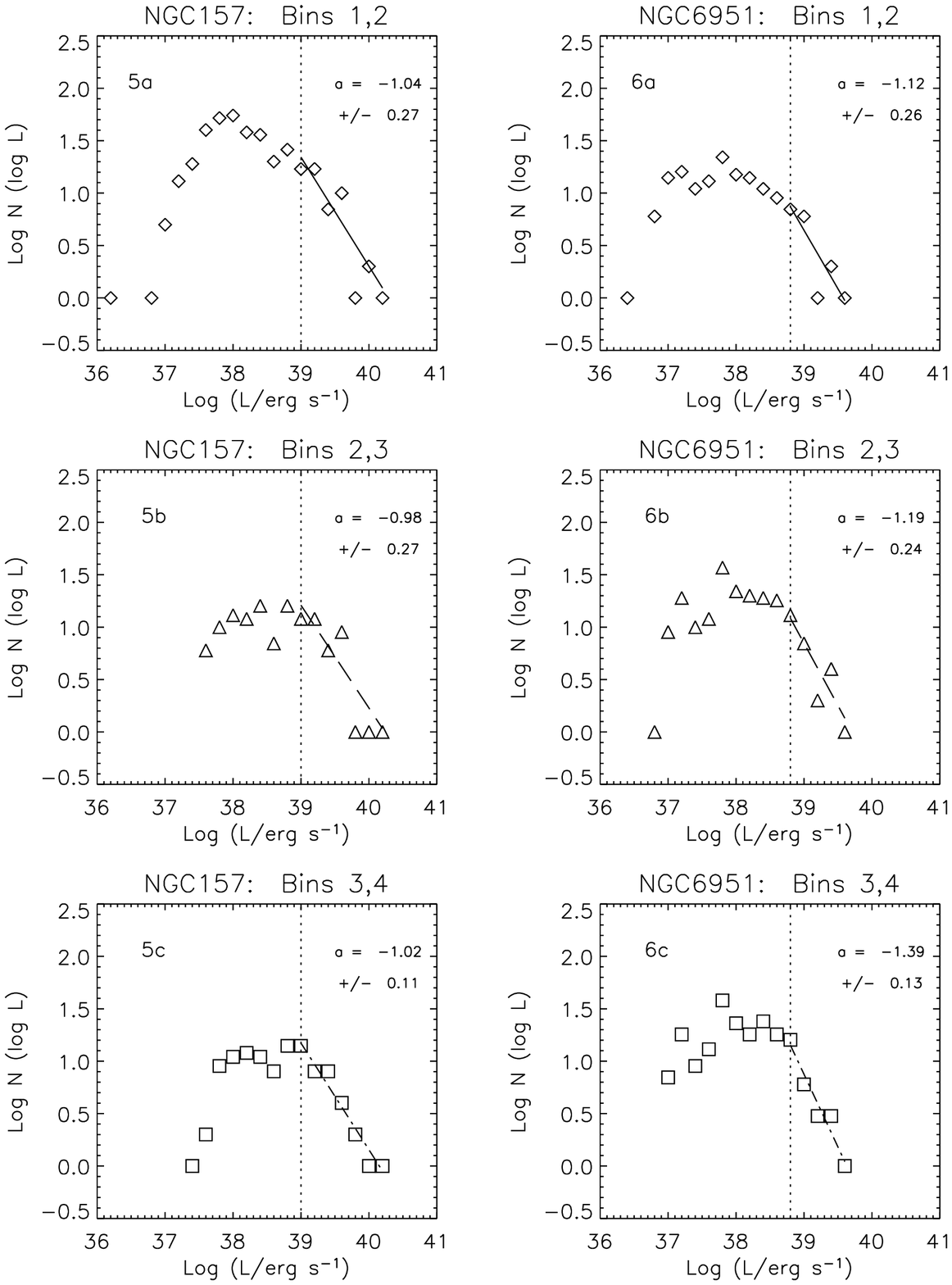}
\end{figure*}
\setcounter{figure}{6}

It is apparent in Figures~\ref{M51hlf} -- 6 that the
location of $\lsat$ is essentially indistinguishable between the
three subsets.  The transition point generally appears to be more
ambiguous in the oldest binnings (panels $c$) as compared to the
youngest (panels $a$), but more galaxies must be examined to determine
whether that pattern is real.  If so, then it would be consistent with
a greater contribution from evolved populations (Paper~I).
The ambiguities again reflect our discussion above, that
the isochrones for these galaxies are insensitive to aging effects,
since the spatial scatter is of order the separation for isochrones
over the nebular evolution time scale.  

%

A number of other factors probably also contribute to the
insensitivity of the isochrones to aging effects.  First, the
models in Paper~I (e.g., Figure~\ref{figmod} above) assumed the
instantaneous creation of all objects in the \hlf, 
whereas triggering by the spiral density wave most likely has a finite
timescale.  Since most \hii\ regions recombine on timescales of
$\lesssim 10$ Myr, we would not expect any significant population to
be visible at 24 Myr.  For the most part, this is consistent with the
fairly negligible \Ha\ emission at the 24 Myr isochrone for most of
the arms in M51, M100, and NGC 157.  NGC 6951 shows spatially more
extensive star formation, with \hii\ regions also seen along the 24
Myr isochrone; this therefore implies an extended timescale for star
formation at any given location.  An extended star-formation timescale 
is also equivalent to a spatial scatter for objects of any given age,
as mentioned above.  For the \hlf s, this will blur the distinction
between the youngest subset and next-to-youngest.  It may also be that
the luminosity fading of the \hii\ regions is even faster than our
extreme model in Paper~I (see Paper~I, Figure~2$f$).  Another factor
that affects the quantitative comparison to the \hlf\ evolution is the
uncertainty in identifying and fitting the locus of the zero-age
density wave.  Finally, our simple analytic description for the
rotation curve and pattern speed are idealized forms that do not
describe the details of the gas kinematics (see \S 4 below), which
also affect the location of the spatial isochrones.  

Thus, our test of the evolutionary features in the \hlf\ for these
galaxies is inconclusive, though an upper limit on the evolutionary
effects is indicated.  For most of the galaxies, the spatial
isochrones do not offer the necessary separation to isolate
populations of \hii\ regions triggered in the arm density waves.

\bigskip
\section{Locating corotation with spatial isochrones}

We now describe our analysis of spatial isochrones, which offer
some interesting insights on spiral arm kinematics.

While measurements of rotation curves are fairly straightforward,
historically it has been much more difficult
to estimate the spiral density wave pattern speed, or equivalently, to
determine the location of the corotation radius.  Several
methods for determining the corotation radius or pattern speed
have been proposed in the past.  The earliest proposal
was based on Lin's (1970) conjecture that organized, two-armed
spiral patterns exist only within the corotation circle, and that arms
become more fragmented outside.  Shu, Stachnik, \& Yost (1971) were
among the first to apply this conjecture to physical galaxies,
placing the corotation where the distribution of
\hii\ regions is seen to end.  Tremaine \& Weinberg (1984)
developed an approach for pattern speed determination 
based on measurement of surface brightness and velocity
distributions of luminosity density tracers, assuming continuity
in the tracer.  Canzian's (1993) method for
determining corotation relies on the qualitative difference
in the residual velocity field inside versus outside
of corotation.  Elmegreen, Elmegreen, \& Seiden (1989) identified
resonances through breaks in spiral arms, and Cepa \& Beckman (1990)
utilized the radial star formation efficiency.  Elmegreen, 
Wilcots \& Pisano (1998) determined corotation from the transition of 
the inward to outward streaming gas velocity.  Finally,
Clemens \& Alexander (2001) used high-resolution
spectroscopic data to find signatures of the pre- and post-shock
gas across the spiral arms; this information determines
the shock velocity, and hence, the pattern speed.  In what follows, we 
explore a new approach that uses the nebular spatial distribution to
constrain the corotation radius.

For a galaxy with a flat rotation curve, the corotation radius $\rco$ is 
related to the angular pattern speed $\Omega_p$ as,
\begin{equation}
\Omega_p = \frac{\vcirc}{r_{\rm co}} \quad ,
\end{equation}
where $\vcirc$ is the circular velocity.
Assuming that star formation is triggered by spiral density waves, the
locus of the youngest \hii\ regions essentially indicates the location
of the density wave.  As the nebulae age, their positions will be
determined by both the spiral arm pattern speed and the galaxy's
rotation curve.  It is therefore possible to determine spatial
isochrones for a given galaxy, that represent the positions of objects
that all formed simultaneously within a density wave.

Our code computes spatial isochrones, given
an initial locus for the spiral arm pattern.  The
\hii\ region positions are deprojected and converted to polar
coordinates, and we fit the spiral pattern at the arm's leading
edge, using a polynomial function.  We then 
compute spatial isochrones from the locus of the spiral arm
pattern, rotation curve, and putative corotation radius.  We assume a
flat rotation curve, except for the innermost regions $r < r_{\rm
circ}$, where we assume $v \propto r$ (Table~\ref{params}).

\begin{figure*}
\vspace*{0.3 truein}
\epsscale{2.0}
\plotone{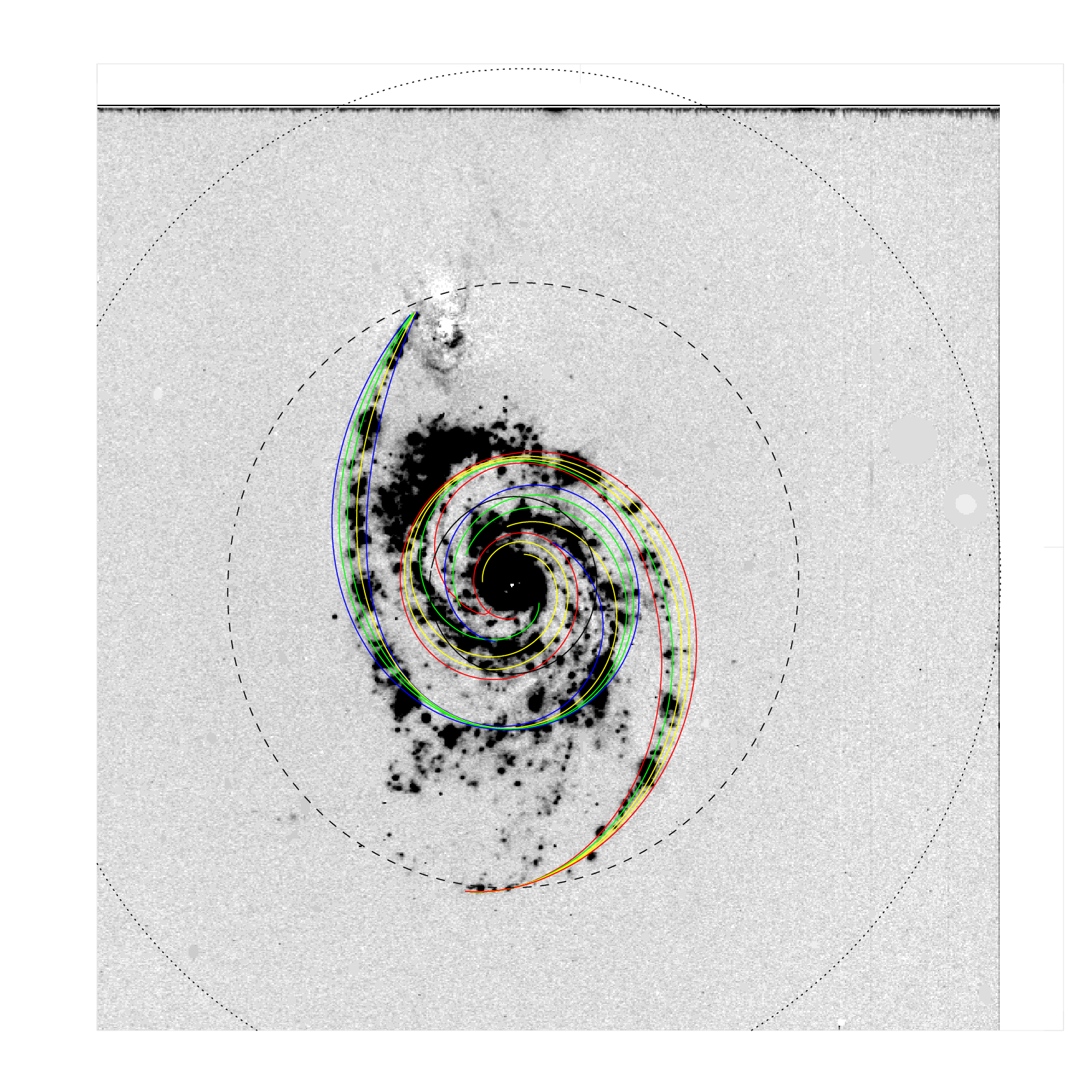}
\caption{\Ha\ images of the sample galaxies, overlaid with
spatial isochrones for $-12,\ 0,\ 12,\ 24,$ and 36 Myr.  These are
respectively shown as red, green, yellow, yellow, red for one arm; and
blue, yellow, green, green, blue for the other.  The black dashed line
shows the adopted corotation radius; the inner and outer Lindblad
resonances are shown by the solid and dotted lines, respectively.
Figure~\ref{isochrones}($a$) shows M51, with \Ha\ data from Rand (1992).
\label{isochrones}}
\end{figure*}

Our code includes a widget to easily adjust the assumed corotation
radius.  Varying $\rco$ by hand, we found a value of
$\rco$ that determines a set of isochrones having a noticeable
correspondence to the spatial configuration of the \hii\ regions.
These values of $\rco$ turn out to correspond to the ends of the
spiral arms in the star-forming disk.  As we discussed above,
other investigators (e.g., Shu et al. 1971; 
Rohlfs 1977; Kenney {\etal}1992; Zhang, Wright, Alexander 1993) have
identified $\rco$ with the ends of the spiral arms in the past.
For three of our four galaxies, Figure~\ref{isochrones} shows that
these values of $\rco$ also yield inner Lindblad resonances (ILRs)
that coincide well with the radial onset of significant star formation
in the arms.  Figure~\ref{isochrones}$b$ also suggests a correspondence
between the outer Lindblad resonance (OLR) and the edge of the star-forming
disk in M100.

\begin{deluxetable}{lccccccl}
\footnotesize
\tablecaption{Galaxy and resonance parameters \label{params}}
\tablewidth{0pt}
\tablehead{
\colhead{Galaxy} & \colhead{Distance} & \colhead{$\vcirc$} 
& \colhead{$r_{\rm circ}$}
& \colhead{$\rco$} & \colhead{ILR} & \colhead{OLR} 
& Reference\tablenotemark{a} \\
& (Mpc) & ($\kms$) & (kpc)
& (kpc) & (kpc) & (kpc) & 
}
\startdata
M51	& \phn 9.6 & 220 & 0.93	& 14 & 4.1 & 24.9 & Rand (1993) \\
M100	& 16.1 & 210 & 1.56 & 12 & 3.5 & 20.5 & Sempere {\etal}(1995)\\
NGC 157	& 22.2 & 170 & 0.54 & \phn 6 & 1.8 & 10.2 & Sempere \& Rozas
(1997); Ryder {\etal}(1998) \\
NGC 6951 & 19.0 & 235 & 0.46 & 10 & 2.9 & 17.1 & Rozas {\etal}(2002) \\
\enddata
\tablenotetext{a}{References for rotation curve parameters:  
$r_{\rm circ}$ is the radius beyond which the circular velocity
is taken to be constant, at value $\vcirc$. }
\end{deluxetable}

For all four galaxies, the regions near the ILRs
display the characteristic ``twin-peaks'' morphology (Kenney
{\etal}1992) or ``twin-peaks-plus-inner-bar'' (Zhang et al. 1993),
together with an annular zone between the twin peaks 
and the inner bar/nucleus that is swept clear of star formation.  This
combination of central morphologies is common among mid- to early-type
galaxies, and is related to the orbital behavior of gas near the
double inner Lindblad resonances that are present in these galaxies
with high central potential wells (Zhang et al. 1993, and references
therein), namely, the inner-inner-Lindblad resonance (IILR) and the
outer-inner-Lindblad resonance (OILR).  This morphology allows the 
effective central channeling of gas and fueling of nuclear activity.
Note that our simplified assumed rotation curves yield only one ILR
which presumably is located between the IILR and OILR
of the actual rotation curves.

For M51 and M100, Figure~\ref{isochrones} shows that setting $\rco$ to
the tips of the spiral arms causes the isochrones to bunch or
converge, but not cross over, at azimuthal locations near
large clumps of nebulae.  These clumps are not confined within the
isochrones, and extend to radii substantially beyond them; however,
their azimuthal coincidence with the isochrone convergences strongly
suggests that these loci are associated with dynamical triggering of
the star formation.  These patterns are most likely related to
non-circular motions and asymmetry in the mass distributions.  Indeed, 
both M51 (see Garc\'\i a-Burillo {\etal}1993) and M100
(Garc\'\i a-Burillo {\etal}1998) show evidence of an inner tri-axial
mass distribution.  
It is also apparent in Figure~\ref{isochrones} that the
spatial isochrones only cover a fraction of the active star-forming
regions in the disks, which again must be caused by non-circular
kinematics.  However, our experimentation with rotation curve
parameters showed that to first order, our simple assumptions for the
rotation curve (flat, with a linear inner portion) worked best in
matching the \Ha\ morphology of the arms.

\setcounter{figure}{6}
\begin{figure*}
\vspace*{0.3 truein}
\epsscale{2.0}
\plotone{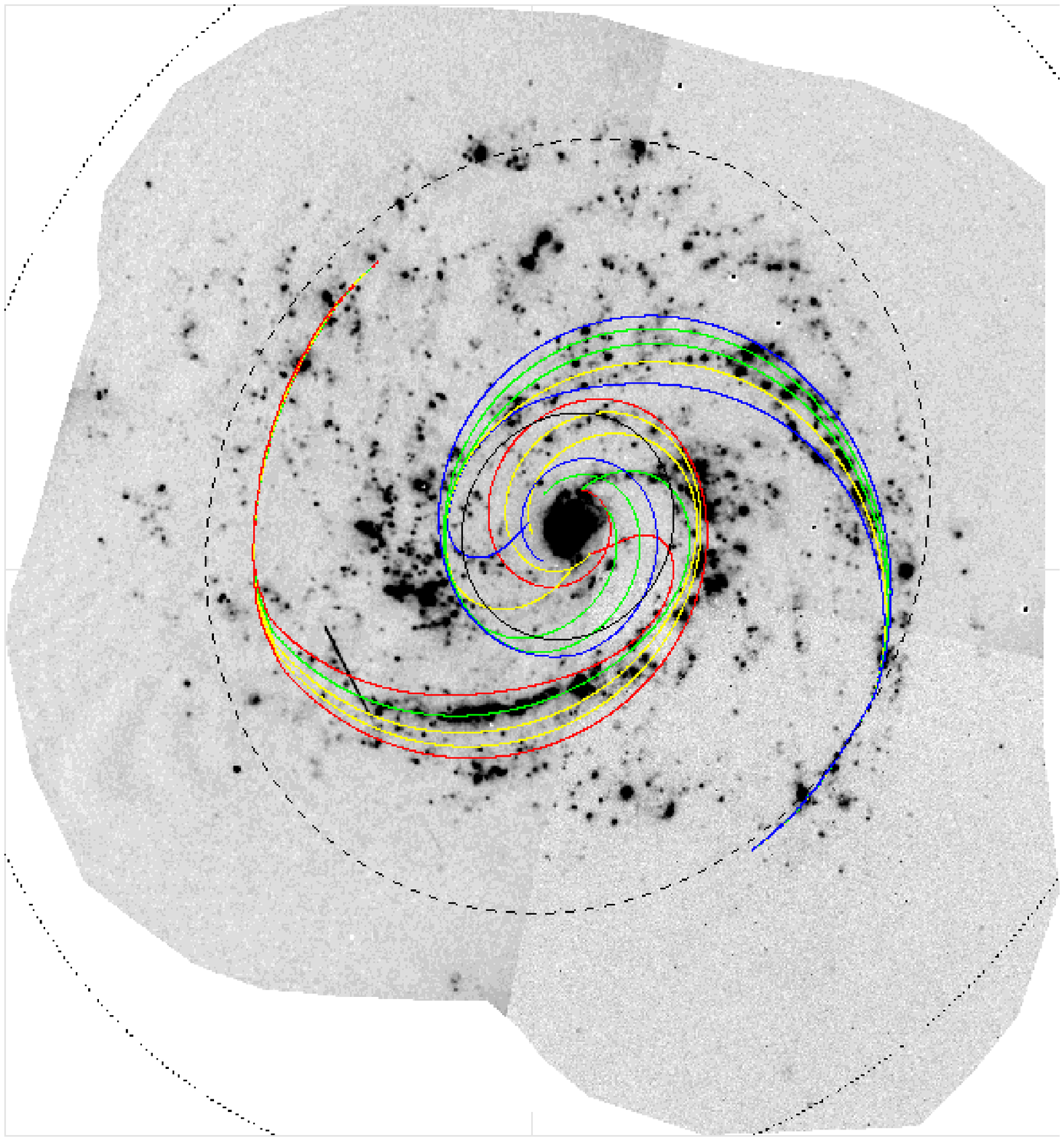}
\caption {($b$)  M100; \Ha\ data from Knapen et al. 1993}
\end{figure*}

\setcounter{figure}{6}
\begin{figure*}
\vspace*{0.3 truein}
\epsscale{2.0}
\plotone{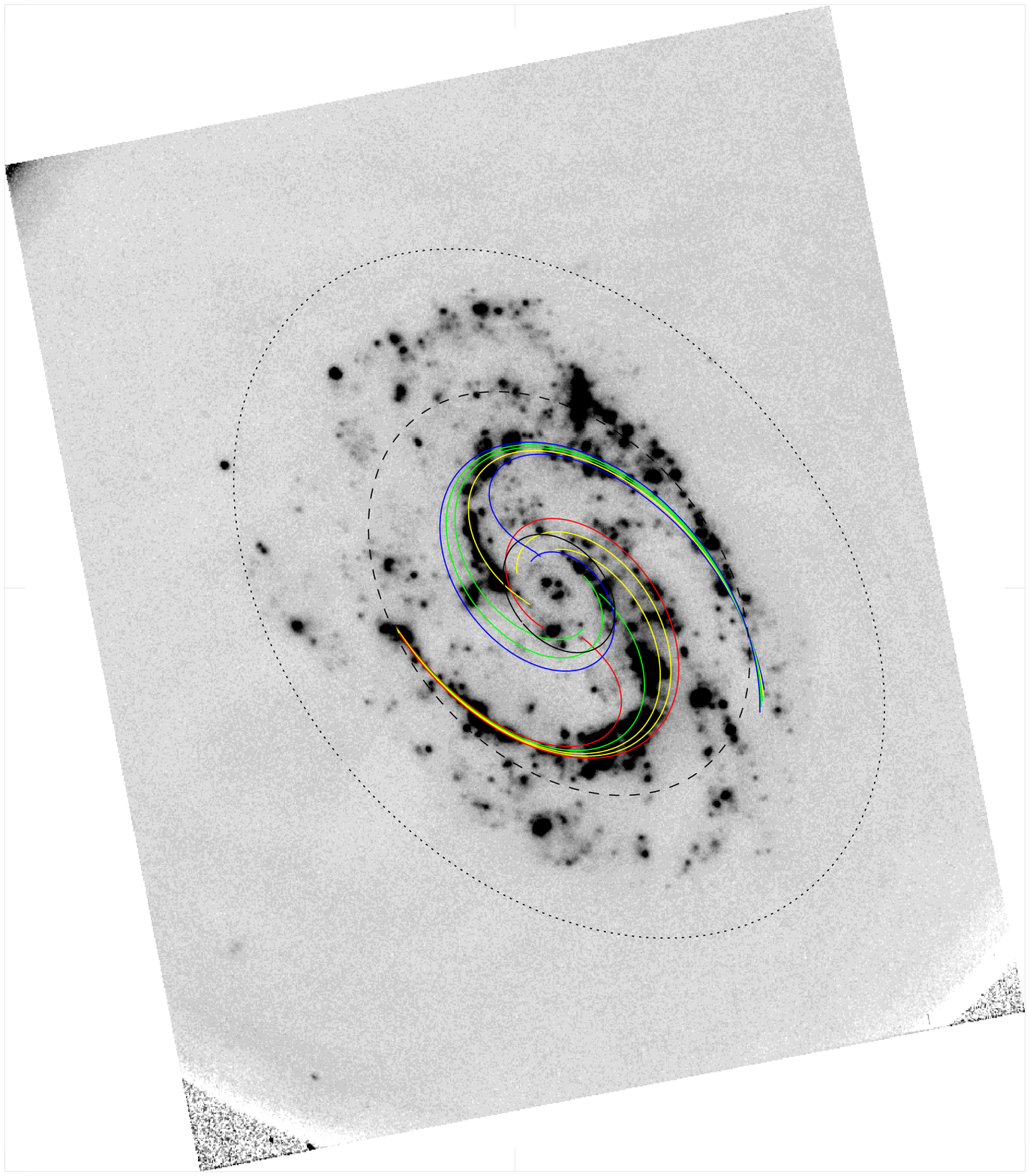}
\caption{ ($c$)  NGC 157; \Ha\ data from Rozas et al. 1996}
\end{figure*}

\setcounter{figure}{6}
\begin{figure*}
\vspace*{0.3 truein}
\epsscale{2.0}
\plotone{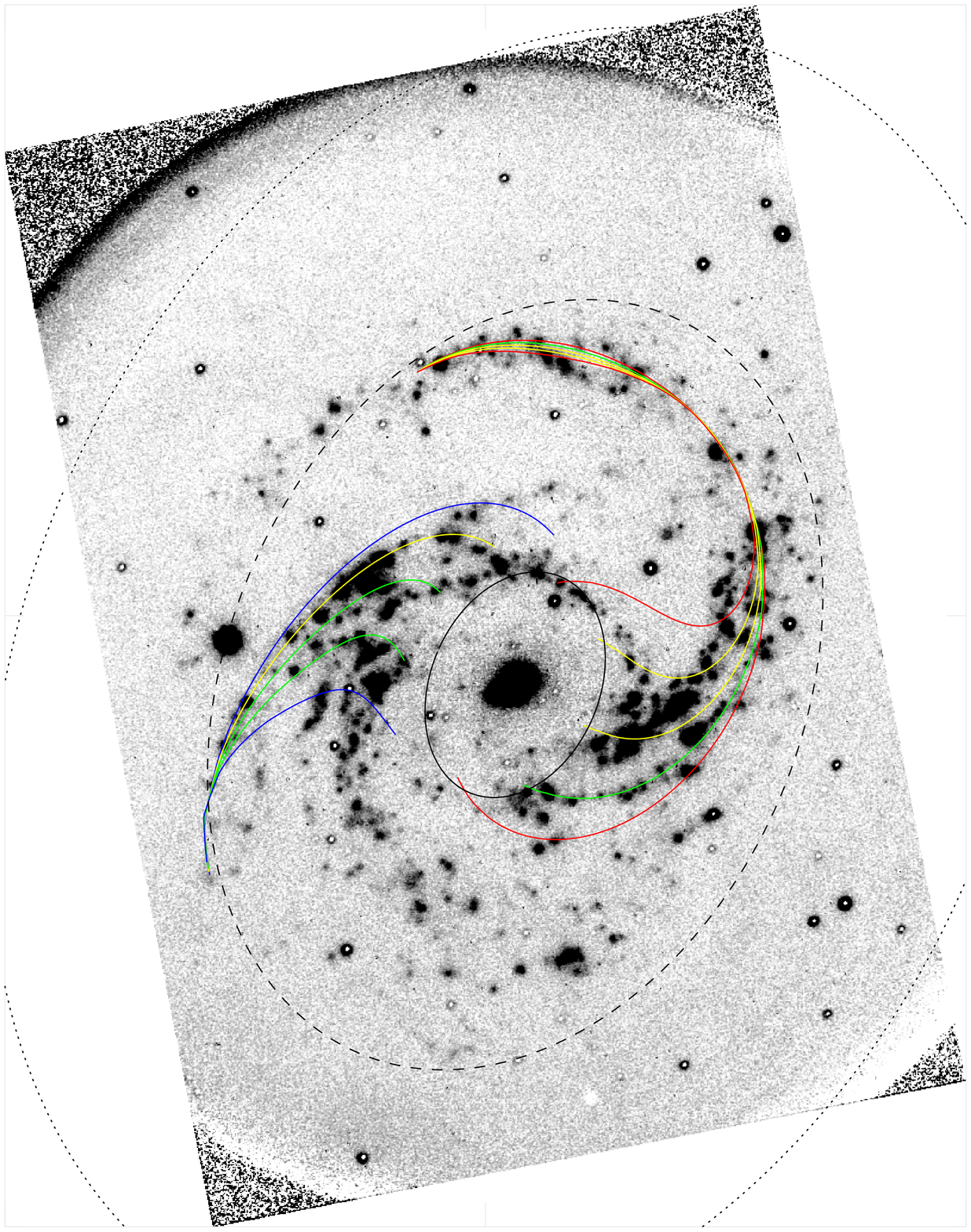}
\caption{ ($d$)  NGC 6951; \Ha\ data from Rozas et al. 1996}
\end{figure*}

\begin{figure*}
\vspace*{0.3 truein}
\epsscale{2.0}
\plotone{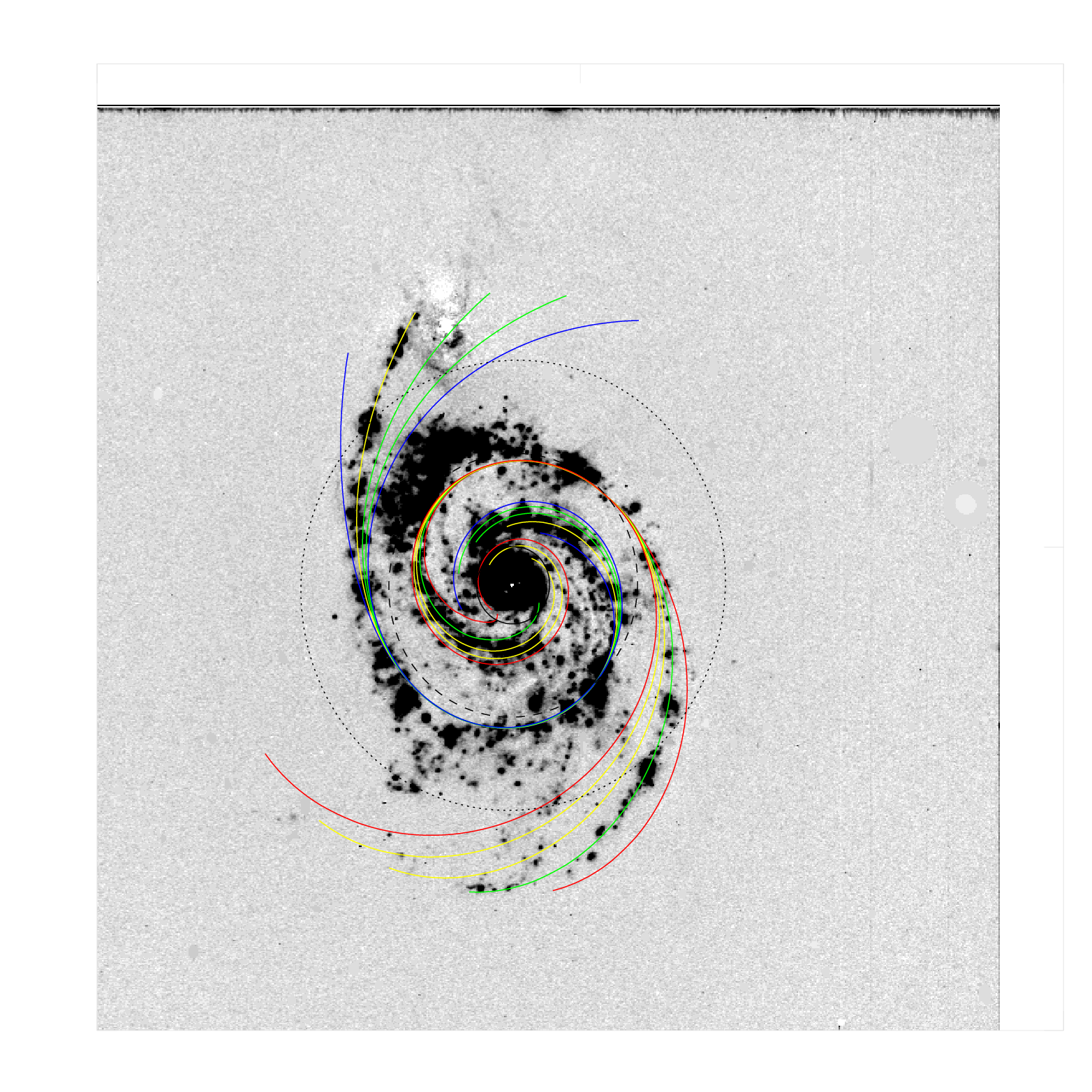}
\caption {Same as Figure~\ref{isochrones}$a$, but with the M51 corotation
at 9.6 kpc suggested by Elmegreen {\etal}1992. 
\label{M51bad}}
\end{figure*}

A counterexample is shown in Figure~\ref{M51bad} for M51.  One previous
estimate for $\rco$ in this galaxy is 9.6 kpc (Elmegreen {\etal}1992).
It is apparent that the isochrones do not correspond to the
\hii\ region positions as cleanly as in Figure~\ref{isochrones}$a$.
The arms show a smooth, continuous morphology in the region around the
suggested $\rco$, whereas the predicted isochrones change character and
diverge. 

Table~\ref{params} shows our assumed rotation curve parameters and our
estimates for the $\rco$, ILR, and OLR.  We now compare our results
for $\rco$ with those in the literature, in particular, estimates
derived from kinematic models.  There are no modeled estimates of
$\rco$ for NGC 6951.

\bigskip\bigskip
\subsubsection{M51}

Our analysis suggests that the corotation radius is twice as large as
most previous determinations for M51.  Both morphological estimates
(e.g., Vogel {\etal}1993; Elmegreen {\etal}1992; Knapen et al. 1992)
and kinematic model 
estimates (Garc\'\i a-Burrillo {\etal}1993b) found $\rco$ of $126\arcsec
- 161\arcsec$, in contrast to our estimate of $301\arcsec$.  We note
that M51 is one of the few examples where the arm definition is narrow enough 
that the $R$-band and \Ha\ or CO observations can clearly distinguish
the stellar arms from the new star-formation (Rand \& Kulkarni 1993;
Garc\'\i a-Burrillo {\etal}1993a).  At corotation, the youngest
regions should switch from the inside edge of the arm to the outside;
yet the data shown by Rand \& Kulkarni (1990) and Garc\'\i a-Burrillo
({\etal}1993a) show no evidence for this transition at the smaller
$\rco$ suggested in the past. 

\subsubsection{M100}

For M100, our corotation radius is larger than previous estimates, but
by a smaller margin.  Estimates by Garc\'\i a-Burrillo {\etal}(1998) and
Canzian \& Allen (1997), based on kinematic modeling, and also Elmegreen
{\etal}(1992) all agree fairly 
well with the range of $82\arcsec - 114\arcsec$ estimated for $\rco$ by
Sempere {\etal}(1995).  Our
estimate of $\rco \sim 154\arcsec$ is larger, but is based on more
simplistic morphological and kinematic assumptions; most of the above
studies consider the more realistic, bar-dominated nature of the
rotation curve.  We also note that the ends of the arms are not
clearly delineated in this galaxy.

\subsubsection{NGC 157}

Our estimate of $\rco\sim 56\arcsec$ agrees well with estimates in
the literature based on detailed kinematic models.  Sempere \& Rozas
(1997) find $\rco\sim 50\arcsec$, and Fridman {\etal}(2001) find
$36\arcsec<\rco < 63\arcsec$ by two different methods.  Elmegreen
{\etal}(1992) also estimate corotation to be at 56$\arcsec$ based on
morphological arguments.

\smallskip

Spatial isochrones thus appear to offer a new technique for
constraining the corotation radius for a first-order approximation.  
We can set a spatial correspondence between:  (1) Corotation and
the end of the nebular spiral arms; (2) Isochrone bunching
and \hii\ region clumps; and possibly (3) Outer Lindblad resonance and
the edge of the star-forming disk.  These physical coincidences 
suggest that the location of corotation has been identified 
to first order.  Thus, while some authors often identified the ends of
the star-forming arms with the corotation resonance, and others with
the OLR, our analysis supports the former.  We also note that in
wide-format, blue photographic images available via the NASA/IPAC
Extragalactic Database (e.g., from the Sandage-Bedke Atlas), there is
no evidence in any of our galaxies that the dust lane crosses over
from the inside to the outside of the arms at any point inside our
adopted $\rco$.

\section{Conclusion}

To better understand the behavior of the \hlf\ and
its ability to probe global star-formation properties, we have 
re-examined its relation to the \hii\ region size distribution, and
investigated its evolution with respect to density wave triggering in 
spiral galaxies.

We have analytically re-evaluated the relationship between the \hlf\ and
the \hii\ region size distribution.  We suggest that {\it the size
distribution is intrinsically described as a power law,} as physically
related to the \hlf, rather than its conventional description by an
exponential relation.  We find that the physical relation between
nebular luminosity and size implies that the slopes of the size
distribution $b$ and the luminosity function $a$ should be related as
$b=2-3a$.  The normally observed value $a=2$ thus yields $b = 4$,
which we find to be consistent with observations.  We find that
the slope transition radius $\log R_{\rm sat}\sim 2.1$ pc
corresponding to the transition luminosity $\log \lsat\sim 38.5$ is
also consistent with observations.

We have developed a code to compute spatial isochrones from the
rotation curve, given the current locus of the spiral density waves.
Our code allows quantitative comparison of the nebular spatial
distribution with the spatial isochrones.  We find that the best
correspondence for the arm \hii\ regions occurs when the corotation
radius coincides with the ends of the star-forming arms.  This value
for $\rco$ also yields a coincidence between the Inner Lindblad
Resonance with the radial onset of star formation in the arms for
three of the four galaxies.  There may also be a hint that the Outer
Lindblad Resonance coincides with the outer edge of the star-forming
disk in at least one galaxy.  {\it The spatial isochrones therefore
appear to offer a new technique for determining the corotation radius 
and principal resonances.}    

We had hoped that these isochrones would isolate subsets of \hii\
regions that could then be used to study evolutionary effects of the
nebular population.  However, for three of the four galaxies, the
isochrones do not yield adequate spatial separation over the nebular
evolution timescale, relative to the spatial scatter at any given
age.  Encouragingly, we do find that the spatial concentration of the
objects along the isochrones is consistent with the overall nebular
evolutionary timescale in these galaxies.
However, the data do
suggest periods of extended star formation at any given location that
blur the diagnostics.  Combined with uncertainties in the kinematic
parameters, this greatly limits the ability to obtain clear quantitative
constraints on \hlf\ evolutionary parameters as related to the density
wave passage.  The observed rate of evolution, however, does set
limits on the importance of this process in producing the observed
arm-interarm variation in the \hlf.

\acknowledgments

We are pleased to acknowledge useful discussions with Ron Allen and
Blaise Canzian, and also comments from the anonymous referee.  Johan
Knapen, Rich Rand, and 
Maite Rozas kindly provided their \Ha\ observations of the galaxies
studied here.  We thank Laura Woodney for tips with IDL.  Much of this
work was carried out by the PI while at the Space Telescope Science
Institute, and by JSP and VJM under the STScI Summer Student Program. 
MSO also acknowledges support from the National Science Foundation, 
grant AST-0204853.  XZ is supported in part by funding from the Office
of Naval Research.  

\vfill\eject


\end{document}